\documentclass[11pt]{article}
\usepackage{color,graphicx}
\usepackage{amssymb,amsmath}
\usepackage{mathrsfs}
\usepackage{bbm}

\usepackage{authblk}

\allowdisplaybreaks \numberwithin{equation}{section}
\newtheorem{thm}{Theorem}[section]
\newtheorem{prp}[thm]{Proposition}
\newtheorem{lem}[thm]{Lemma}
\newtheorem{dfn}[thm]{Definition}
\newenvironment{defn}{\begin{dfn} \rm }{\end{dfn}}

\newtheorem{example}[thm]{Example}
\newenvironment{exa}{\begin{example} \rm }{ \end{example}}
\newtheorem{remark}[thm]{Remark}

\newenvironment{rmk}{\begin{remark} \rm }{\hfill $\Box$ \end{remark}}
\newenvironment{prf}{\noindent {\it Proof:} \ }{\hfill $\Box$}

\newcommand\od{\mathrm{d}}
\newcommand\p{\partial}

\newcommand{\nn}{\nonumber}

\newcommand{\al}{\alpha}

\newcommand{\om}{\omega} \newcommand{\Om}{\Omega}
\newcommand{\Gm}{\Gamma}

\newcommand{\dt}{\delta}

\newcommand{\vp}{\varphi}

\newcommand{\res}{\mathrm{res}}

\newcommand\fg{\mathfrak{g}}
\newcommand\Z{\mathbb{Z}}
\newcommand\C{\mathbb{C}}

\newcommand\cA{\mathcal{A}}
\newcommand\cD{\mathcal{D}}

\newcommand\cG{\mathcal{G}}
\newcommand\cH{\mathcal{H}}

\newcommand\cM{\mathcal{M}}

\newcommand\cP{\mathcal{P}}
\newcommand\cU{\mathcal{U}}
\newcommand\cV{\mathcal{V}}

\newcommand\fD{\mathfrak{D}}

\newcommand{\set}[1]{\left\{#1\right\}}
\newcommand\ra{\rangle}
\newcommand\la{\langle}

\newcommand{\bm}[1]{\mathbf{#1}}
\newcommand{\bt}{\mathbf{t}}

\begin{document}

\title{An Extension of the Kadomtsev--Petviashvili Hierarchy
and its Hamiltonian Structures}

\author[1]{Chao-Zhong Wu}
\author[2]{Xu Zhou}
\affil[1]{School of Mathematics and Computational Science, Sun
Yat-Sen University,

Guangzhou 510275, P.R. China. wuchaozhong@sysu.edu.cn }

\affil[2]{Department of Mathematical Sciences, Tsinghua University,
Beijing 100084, P.R. China. x-zhou09@mails.tsinghua.edu.cn }

\date{}
\maketitle

\begin{abstract}
In this note we consider a two-component extension of the
Kadomtsev--Petviashvili (KP)
 hierarchy represented with two types of pseudo-differential operators, and
 construct its Hamiltonian structures by using the $R$-matrix formalism.
\end{abstract}

\parskip 1ex

\section{Introduction }

The Kadomtsev--Petviashvili hierarchy plays a fundamental role in
the theory of integrable systems. There are several ways to define
the KP hierarchy, and one contracted way is via Lax equations of
pseudo-differential operators as follows. Let
\begin{equation} \label{Lckp}
L_{KP}=\p+u_1\p^{-1}+u_2\p^{-2}+\dots, \quad \p=\frac{\od}{\od x},
\end{equation}
be a pseudo-differential operator with scalar coefficients $u_i$
depending on the spacial coordinate $x$. The KP hierarchy is
composed by the following evolutionary equations
\begin{equation}\label{kp}
 \frac{\p L_{KP}}{\p t_k}=[(L_{KP}^k)_+, L_{KP}],\quad
 k=1,2,3,\dots.
\end{equation}
Here the subscript ``$+$'' means to take the differential part of a
pseudo-differential operator. The hierarchy \eqref{kp} is known to
possess a series of bi-Hamiltonian structures, which can be
constructed, for instance, by the $R$-matrix formalism \cite{STS}.

The KP hierarchy \eqref{kp} has been generalized to multicomponent
versions with scalar pseudo-differential operators replaced by
matrix-value ones \cite{DJKM, BtK, KvdL}. In such generalizations,
the pseudo-differential operators are required to admit certain
extra constraints, and it is probably why no Hamiltonian structures
underlying have been found. Towards overcoming this difficulty, a
step was made by Carlet and Manas \cite{CM}, who solved the
constraints in the $2$-component case and parameterized the matrix
operators with a set of ``free'' dependent variables. The study of
the $2$-component KP hierarchy was also motivated by the development
of infinite-dimensional Frobenius manifolds in recent years,
especially those associated with the bi-Hamiltonian structures for
the Toda lattice and the $2$-component BKP hierarchies \cite{CDM,
WX, WZ}.

Inspired by the Toda lattice hierarchy, we now consider an extension
of the KP hierarchy from the viewpoint of scalar pseudo-differential
operators rather than matrix-value ones. More exactly, given a pair
of scalar operators
\begin{equation} \label{PhP}
P= D+\sum_{i\ge1}u_i  D^{-i}, \quad \hat{P}=
D^{-1}\hat{u}_{-1}+\sum_{i\ge0}\hat{u}_i D^i
\end{equation}
with $D$ being a ``derivation'' on some differential algebra that
contains functions $u_i$ and $\hat{u}_i$, we want to define the
following commutative flows:
\begin{align}\label{ekp1}
& \frac{\p }{\p t_k}(P,\hat{P})=\left([(P^k)_+, P],[(P^k)_+,
\hat{P}]\right), \\
& \frac{\p}{\p \hat{t}_k}(P,\hat{P})=\left([-(\hat{P}^k)_-, P],
[-(\hat{P}^k)_-, \hat{P}]\right) \label{ekp2}
\end{align}
for $k=1, 2, 3, \dots$. It will be seen that if one takes $D=\p-\vp$
with $\vp$ being an unknown function of $x$, then the hierarchy
\eqref{ekp1}--\eqref{ekp2} is well defined (see
Section~\ref{sec-eKP}). Such kind of hierarchies appeared in the
work \cite{SB} of Szablikowski and Blaszak as a dispersive
counterpart of the Whitham hierarchy (see, for example, \cite{TT}).
Their version in fact involves $N$ operators of the form $\hat{P}$
with $D$ replaced by $\p-\vp_i$ for distinct functions $\vp_i$
($1\le i\le N$). However, the convergence property of the operators
$\hat{P}^k$, which can contain infinitely many positive powers in
$D$, seems not have been taken into account before. In this note, we
will only consider the case $N=1$, and illustrate that the operators
$\hat{P}^k$ as the so-called pseudo-differential operators of the
second type introduced by us \cite{LWZ} in recent years. Such kind
of operators converge according to a suitable topology. As to be
seen, the extended KP hierarchy \eqref{ekp1}--\eqref{ekp2} can be
reduced to the $2$-component BKP hierarchy \cite{DJKM, LWZ} and the
constrained KP hierarchy (see, for example, \cite{ANP, BX, Kr})
under suitable constraints.

Observe that the flows \eqref{ekp1}--\eqref{ekp2} are defined on a
subset of the Lie algebra $\cG^-\times\cG^+$, with $\cG^\mp$ being
the algebras of pseudo-differential operators of the first and the
second types respectively. It is natural to apply the $R$-matrix
scheme to search for Hamiltonian structures underlying
\eqref{ekp1}--\eqref{ekp2}, such as what we have done for the Toda
lattice and the $2$-component BKP hierarchies \cite{Wu} (cf.
\cite{Ca, WX2}). It will be seen that the simple but useful
$R$-matrix found in \cite{Wu} is also feasible in the current case,
so a series of bi-Hamiltonian structures for the hierarchy
\eqref{ekp1}--\eqref{ekp2} will be derived. Furthermore, such
bi-Hamiltonian structures can be naturally reduced to that for the
constrained KP hierarchy. Note that the bi-Hamiltonian structure for
the constrained KP hierarchy used to be obtained by Oevel and
Strampp \cite{OS}, Cheng \cite{CY} and Dickey \cite{Di},
respectively, with different methods. In a recent paper \cite{LZZ},
the central invariants for the bi-Hamiltonian structure have been
calculated, which shows that the constrained KP hierarchy is the
so-called topological deformation of its dispersionless limit.

This article is organized as follows. In the next section we recall
the notions of pseudo-differential operators of the first and the
second type, and then check in detail that the extended KP hierarchy
\eqref{ekp1}--\eqref{ekp2} is well defined. In Section~3, we review
briefly the $R$-matrix method for deriving Hamiltonian structures,
and apply this method to the hierarchy \eqref{ekp1}--\eqref{ekp2}
and its reductions. Finally some remarks will be given.

\section{An extension of the KP hierarchy}\label{sec-eKP}

\subsection{Preliminary notations}
Let $\cA$ be a commutative associative algebra, and $\p: \cA\to\cA$
be a derivation. The set of pseudo-differential operators is
\[
\cA((\p^{-1}))=\set{\sum_{i\le k}f_i\p^i\mid f_i\in\cA, k\in\Z},
\]
in which the product is defined by
\begin{equation}\label{pro}
f \p^i\cdot g \p^j=\sum_{r\geq0}\binom{i}{r}f\, \p^r(g)\,
\p^{i+j-r}, \quad f,\,g\in\cA.
\end{equation}
Clearly one has the commutator $[\p, f]=\p(f)$ for any $f\in\cA$.

From now on we assume the algebra $\cA$ to be a graded one. Namely,
$\mathcal{A}=\prod_{i\geq0}\cA_i$, such that
\[
\cA_i\cdot\cA_j\subset\cA_{i+j}, \quad \p(\cA_i)\subset\cA_{i+1}.
\]
Denote $\cD^-=\cA((\p^{-1}))$, which is called the algebra of
pseudo-differential operators of the first type over $\cA$. In
comparison, by the algebra of pseudo-differential operators of the
second type over $\cA$ it means
\begin{equation}\label{Dpl}
\cD^+=\left\{ \sum_{i\in\Z}\sum_{j\ge \max\{0,k-i\}}a_{i,j} \p^i
\mid a_{i,j}\in\cA_j, k\in\Z \right\},
\end{equation}
which is endowed a product defined also by \eqref{pro}. Observe that
an operator in $\cD^+$ may contain infinitely many positive powers
in $\p$, with appropriate constraints to the degrees of coefficients
as in \eqref{Dpl}, see \cite{LWZ} for details.

Given an arbitrary element $\vp\in\cA_1$ of degree $1$, let us
consider the following two maps defined by replacing $\p$ with
$\p-\vp$, say
\begin{align}\label{}
S_\vp: \quad \cD^\mp&\to\cD^\mp, \nn\\
 \sum {f}_i \p^i &\mapsto \sum {f}_i (\p-\vp)^i.
\label{Svp}
\end{align}
Here we use the same name $S_\vp$ to simplify notations.
\begin{lem}
The map $S_\vp$ is an automorphism on each $\cD^\mp$.
\end{lem}
\begin{prf}
Clearly, on each $\cD^\mp$ the map in \eqref{Svp} is well defined.
In particular,
\begin{align}\label{}
S_\vp(\p^{-1})=&(\p-\vp)^{-1}=\p^{-1}(1-\vp\p^{-1})^{-1} \nn\\
=&\p^{-1}+\p^{-1}\vp\p^{-1} +\p^{-1}\vp\p^{-1}\vp\p^{-1}+\dots.
\end{align}

Note $[\p-\vp,f]=\p(f)$ for any $f\in\cA$, then one has
\[
f (\p-\vp)^i\cdot g (\p-\vp)^j=\sum_{r\geq0}\binom{i}{r}f\,
\p^r(g)\, (\p-\vp)^{i+j-r}, \quad f,\,g\in\cA.
\]
This implies that $S_\vp$ is an endomorphism.

Finally, it is easy to see that $S_\vp$ is injective, and it remains
to show that $S_\vp$ is surjective. This follows from the facts
\begin{align*}
\p=&(\p-\vp)+\vp, \\
\p^{-1}=&(\p-\vp+\vp)^{-1}
\\
=&(\p-\vp)^{-1}-(\p-\vp)^{-1}\vp(\p-\vp)^{-1}
+(\p-\vp)^{-1}\vp(\p-\vp)^{-1}\vp(\p-\vp)^{-1}-\dots.
\end{align*}
The lemma is proved.
\end{prf}

The lemma implies that the algebras $\cD^\mp$ can be represented as
follows:
\begin{align}\label{Dvpm}
\cD^-=&\set{\sum_{i\le k}g_i(\p-\vp)^i\mid g_i\in\cA, k\in\Z},\\
 \cD^+=&\left\{
\sum_{i\in\Z}\sum_{j\ge \max\{0,k-i\}}b_{i,j} (\p-\vp)^i \mid
b_{i,j}\in\cA_j, k\in\Z \right\}. \label{Dvpp}
\end{align}
\begin{dfn}
For a pseudo-differential operator, its $\vp$-expansion is of the
form as elements in \eqref{Dvpm}--\eqref{Dvpp}.
\end{dfn}

Given an operator $A=\sum_{i\in\Z} {f}_i \p^i\in\cD^\mp$, recall
that its differential part is
\begin{equation}\label{}
A_+=\sum_{i\ge0} f_i \p^i,
\end{equation}
while $A_-=A-A_+$ is the negative part, and the residue of $A$ means
\begin{equation}\label{}
\res A=f_{-1}.
\end{equation}
We take the $\vp$-expansion of $A$, say, $A=\sum_{i\in\Z} {g}_i
(\p-\vp)^i$ (in particular, $g_{-1}=f_{-1}$). Then one easily sees
\begin{equation}\label{}
A_+=\sum_{i\ge0} g_i (\p-\vp)^i, \quad A_-=\sum_{i<0} g_i
(\p-\vp)^i, \quad \res A=g_{-1}.
\end{equation}

To prepare for statements below, let us also recall the following
anti-automorphism on each $\cD^\mp$ defined by
\[
\p^*=-\p, \quad f^*=f \hbox{ with } f\in\cA.
\]
Note that such an anti-automorphism is indeed an involution.

\subsection{The extended KP hierarchy}

From now on we take the graded algebra $\cA=\prod_{i\geq0}\cA_i$ to
be the set of formal differential polynomials of certain smooth
function of $x$ parameterizing the loop $S^1$, on which there is
naturally a derivation $\p=\od/\od x$. In our consideration, we can
choose $\cA$ to be generated by a collection of unknown functions
$u_1, u_2, u_3, \dots, \rho, \hat{u}_0, \hat{u}_1, \hat{u}_2, \dots$
of degree $0$ together with an extra function $\vp$ of degree $1$
(the function $\vp$ could be understood as some differential
polynomial of degree 1 in other unknown functions, which are not
necessary in our construction).

Over $\cA$ one has the algebras $\cD^\mp$ of pseudo-differential
operators of the first and the second types respectively. Consider
operators as follows:
\begin{align}\label{P}
P=&(\p-\vp)+\vp+\sum_{i\ge1}{u}_i(\p-\vp)^{-i}\in\cD^-, \\
 \label{Ph}
\hat{P}=&(\p-\vp)^{-1}\rho+\sum_{i\ge0}\hat{u}_i(\p-\vp)^{i}\in\cD^+.
\end{align}

\begin{defn}
The extended KP hierarchy means a system of evolutionary equations
of the unknown functions $u_i, \rho, \hat{u}_i$ given by
\begin{align}\label{Ltth}
& \frac{\p P}{\p t_k}=[(P^k)_+, P], \quad  \frac{\p P}{\p \hat{t}_k}=[-(\hat{P}^k)_-, P],  \\
& \frac{\p \hat{P}}{\p t_k}=[(P^k)_+, \hat{P}],\quad \frac{\p
\hat{P}}{\p \hat{t}_k}=[-(\hat{P}^k)_-, \hat{P}], \label{Lhtth}
\end{align}
where $k=1,2,3,\dots$.
\end{defn}

\begin{rmk}
The first equation in line \eqref{Ltth} gives nothing but the KP
hierarchy. Indeed, the operator $P$ can be recast to the form as
\eqref{Lckp}:
\begin{equation}\label{}
P=\p+\sum_{i\ge1}\tilde{u}_i\p^{-i},
\end{equation}
where $\tilde{u}_i- u_i\in \prod_{i\ge1}\cA_i$. This is partially
why we use the name extended KP hierarchies here, until a more
appropriate name appears. Here we note that the function $\vp$,
arising in the $\vp$-expansion for $P$, is not an independent
unknown function in the KP hierarchy. Another reason for the name is
that we would like to distinguish the hierarchy
\eqref{Ltth}--\eqref{Lhtth} with the so called $2$-component KP
hierarchy \cite{BtK, CM,DJKM, KvdL} represented by matrix-value
pseudo-differential operators (by now we do not see any connection
between them).
\end{rmk}

\begin{rmk}
In \cite{SB}  Szablikowski and Blaszak presented Lax operators in
the form
\begin{align}\label{}
L_\infty=&\p+\sum_{l\ge1}u_{\infty,l}\p^{-l},  \\
L_i=&u_{i,-1}(\p-\vp_i)^{-1}+\sum_{l\ge1}u_{i,l}(\p-\vp_i)^{l},
\quad i=1,2,\dots,N.
\end{align}
They defined a set of auxiliary linear equations
\begin{equation}\label{}
L_\al\Psi_\al=E_\al\Psi_\al, \quad \frac{\p\Psi_\al}{\p
t_{\beta\,n}}=\Om_{\beta\,n}\Psi_\al,
\end{equation}
for some eigenfunctions $\Psi_\al$ and eigenvalues $E_\al$, where
$\al, \beta$ run over $\{\infty, 1, 2, \dots, N\}$ and
\begin{equation}\label{}
\Om_{\al\,n}=\left\{\begin{array}{cl}
                      (L_\infty^n)_{(\infty,+)}, \quad &  \al=\infty;
                      \\
                      \\
                      -(L_i^n)_{(i,-)}, &  \al=i.
       \end{array}\right.
\end{equation}
Here the projections ``$\pm$'' mean as before. The compatibility
conditions between these linear equations are regarded to be the
``quantum'' counterparts of the Whitham hierarchy (without the
logarithm part). In this way, when $N=1$ one gets a hierarchy of the
form \eqref{Ltth}--\eqref{Lhtth} (up to a nonessential
transformation $L_1\mapsto L_1^*$). However, the authors of
\cite{SB} noted that the operators $\Om_{i,n}$ in their construction
do not have a ``limited'' or ``compact'' form. What is more,
Hamiltonian structures for such counterparts seem not have been
studied in the literature.
\end{rmk}

Let us proceed to check that the equations
\eqref{Ltth}--\eqref{Lhtth} are well defined (cf. \cite{SB}). First,
it can be seen that the left and the right hand sides of equations
in line \eqref{Ltth} are operators with only negative part, which
implies that such equations are well defined. Second, the left hand
sides of equations in line \eqref{Lhtth} take a form as
\begin{align}\label{flowminus}
(\p-\vp)^{-1}\frac{\p\vp}{\p\dot{t}_k}(\p-\vp)^{-1}\rho
+(\p-\vp)^{-1}\frac{\p\rho}{\p\dot{t}_k}+\hbox{ differential part},
\end{align}
where $\dot{t}_k$ stand for $t_k$ or $\hat{t}_k$. In order to show
that the right hand sides of the Lax equations in line \eqref{Lhtth}
take a similar form as \eqref{flowminus}, we need to do some
preparation first.

Given any element $f\in\cA$ and integer $k>0$, one has
\begin{align}
f(\p-\vp)^k=&\left((-\p-\vp)^k f\right)^* \nn\\
=&\left(\sum_{r=0}^k\binom{k}{r}(-\p)^r(f)\cdot(-\p-\vp)^{k-r}
\right)^* \nn\\
=&\sum_{r=0}^k\binom{k}{r}(\p-\vp)^{k-r}(-\p)^r(f), \label{fdk}
\end{align}
and
\begin{align}\label{dinf}
(\p-\vp)^{-1}f=f(\p-\vp)^{-1}-(\p-\vp)^{-1}\cdot\p(f)\cdot(\p-\vp)^{-1}.
\end{align}
Suppose that $Q$ is a differential operator, say
\[
Q=\sum_{i\ge0}f_i(\p-\vp)^{i}\in\cD^\mp, \quad f_i\in\cA.
\]
We have
\begin{align}
[Q,\hat{P}]_-=&[Q,(\p-\vp)^{-1}\rho]_- \nn\\
=&f_0(\p-\vp)^{-1}\rho-
\left((\p-\vp)^{-1}\rho\sum_{i\ge0}f_i(\p-\vp)^{i}\right)_- \nn\\
=&f_0(\p-\vp)^{-1}\rho-
\left((\p-\vp)^{-1}\sum_{i\ge0}\sum_{r=0}^i(-1)^r\binom{i}{r}(\p-\vp)^{i-r}\p^r(\rho f_i)
\right)_- \nn\\
=&f_0(\p-\vp)^{-1}\rho- (\p-\vp)^{-1}\sum_{i\ge0}(-1)^i\p^i(\rho
f_i)  \nn\\
=&\left(f_0(\p-\vp)^{-1}-(\p-\vp)^{-1}f_0\right)\rho+
(\p-\vp)^{-1}\sum_{i\ge1}(-1)^{i-1}\p^i(\rho f_{i})  \nn\\
=&(\p-\vp)^{-1}\cdot\p(f_0)\cdot(\p-\vp)^{-1}\rho+
(\p-\vp)^{-1}\sum_{i\ge1}(-1)^{i-1}\p^i(\rho f_{i}), \label{QPm}
\end{align}
of which the third and the last equalities are due to the formulae
\eqref{fdk} and \eqref{dinf} respectively. Observe that the form of
$[Q, \hat{P}]_-$ is consistent with the negative parts in
\eqref{flowminus}, which implies that equations in line
\eqref{Lhtth} are indeed well defined.

In particular, comparing \eqref{flowminus} with \eqref{QPm} one has
\begin{align}\label{phit}
\frac{\p\vp}{\p\dot{t}_k}=&\p\left(\res\,\dot{P}^k(\p-\vp)^{-1}\right),
\\
\frac{\p\rho}{\p\dot{t}_k}=&\sum_{i\ge1}(-1)^{i-1}\p^i\left(\rho\,\res\,
\dot{P}^k(\p-\vp)^{-i-1}\right), \label{rhot}
\end{align}
where $\dot{t}_k=t_k, \hat{t}_k$ and $\dot{P}=P, \hat{P}$. The right
hand side of equation \eqref{phit} is a total derivative, which also
implies that the assumption $\vp\in\cA_1$ in the beginning is
reasonable.

Finally, one can check that the equations in
\eqref{Ltth}--\eqref{Lhtth} are compatible, that is,
$\p/\p\dot{t}_k$ and $\p/\p\dot{t}_l$ are commutative. Therefore
equations \eqref{Ltth}--\eqref{Lhtth} compose an integrable
hierarchy indeed.

\begin{rmk}
Letting $\vp\to 0$, for the flows $\p/\p t_k$ and $\p/\p\hat{t}_k$
to make sense it is necessary to require
\[
\res\,P^k \p^{-1}=\res\,\hat{P}^k \p^{-1}=0.
\]
A nontrivial assumption that fulfills this condition is that $k=1,3,
5, 7, \dots$ and
\[
P^*=-\p P \p^{-1}, \quad \hat{P}^*=-\p \hat{P} \p^{-1}.
\]
In this way what we obtain is nothing but the Lax representation of
the $2$-component BKP hierarchy, see \cite{LWZ} and references
therein.
\end{rmk}

\section{Hamiltonian structures for the extended KP hierarchy}

For the extended KP hierarchy \eqref{Ltth}--\eqref{Lhtth}, we want
to construct its Hamiltonian structures with the $R$-matrix method,
which is similar with the algorithm for the Toda lattice hierarchy
\cite{Wu} (see also \cite{Ca}).

\subsection{$R$-matrix and Poisson brackets}

Let us recall briefly the $R$-matrix formalism and some relevant
results, based on the work \cite{ LP, OR, STS}.

Let $\fg$ be a complex Lie algebra. A linear transformation $R:
\fg\to\fg$ is called an $R$-matrix if a Lie bracket is defined by
\begin{equation}\label{}
[X,Y]_R=[R(X), Y]+[X, R(Y)],\quad X, Y\in\fg.
\end{equation}
For a linear transformation $R$ being an $R$-matrix, a sufficient
condition is that $R$ solves the so-called modified Yang-Baxter
equation:
\begin{equation}\label{YBeq}
[R(X), R(Y)]-R([X, Y]_R)=-[X,Y], \quad X, Y\in\fg.
\end{equation}

Assume $\fg$ to be an associative algebra, whose Lie bracket is
defined by the commutator. We also assume that there is a function
$\la~\ra:\fg\to\C$ that induces a non-degenerate symmetric invariant
bilinear form (inner product) $\la~ ,\,\ra$ as
\[
\langle X, Y\rangle=\la X Y\ra=\la Y X\ra,\quad X, Y\in\fg.
\]
Via this inner product $\fg$  can be identified with its dual space
$\fg^*$. Let $T\fg$ and $T^*\fg$ be the tangent and the cotangent
bundles of $\fg$, with fibers $T_A\fg=\fg$ and $T^*_A\fg=\fg^*$
respectively at any point $A\in\fg$.

Given an $R$-matrix $R$ on $\fg$, for any $f, g\in C^\infty(\fg)$
the following brackets are defiened:
\begin{align} \label{linbr}
\{f, g\}_1(A)&=\frac1{2}\big(\la[A,\od f],R(\od
g)\ra-\la[A,\od g],R(\od f)\ra\big), \\
 \{f, g\}_2(A)&=\frac1{4}\big(\la[A,\od f],R(A\cdot\od
g+\od g\cdot A)\ra-\la[A,\od g],R(A\cdot\od f+\od f\cdot
A)\ra\big), \label{quabr} \\
\{f, g\}_3(A)&=\frac1{2}\big(\la[A,\od f],R(A\cdot\od g\cdot
A)\ra-\la[A,\od g],R(A\cdot\od f\cdot A)\ra\big), \label{cubbr}
\end{align}
where $\od f, \od g \in T_A^*\fg$ are the gradients of $f, g$ at
$A\in\fg$ respectively. Following \cite{LP, OR}, the brackets
\eqref{linbr}--\eqref{cubbr} are called the linear, the quadratic
and the cubic brackets respectively.

\begin{thm}[\cite{LP, OR, STS}]\label{thm-3br}
\begin{itemize}
  \item[(1)] For any $R$-matrix $R$ the linear bracket is a Poisson
bracket.
\item[(2)] Let $R^*$ be the adjoint transformation of $R$ with respect to
the inner product on $\fg$. If both $R$ and its anti-symmetric part
$R_a=\frac1{2}(R-R^*)$ solve the modified Yang-Baxter equation
\eqref{YBeq}, then the quadratic bracket is a Poisson bracket.
\item[(3)]
 If $R$ satisfies the modified Yang-Baxter equation \eqref{YBeq} then
the cubic bracket is a Poisson bracket.
\end{itemize}
Moreover, these three Poisson brackets are compatible whenever all
the above conditions are fulfilled.
\end{thm}

In the present note, the quadratic bracket is the one to be applied,
see below.

\subsection{Poisson bracket on a coupled Lie algebra}

Recall the algebras $\cD^\pm$ of pseudo-differential operators of
the first and the second types over $\cA$. As in \cite{Wu} let us
consider
\begin{equation}
\fD=\cD^-\times\cD^+,
\end{equation}
on which the product is defined diagonally as
\[
(X,\hat{X})\cdot(Y,\hat{Y})=(X Y, \hat{X}\hat{Y}), \quad
(X,\hat{X}),(Y,\hat{Y})\in\fD.
\]
The algebra $\fD$ is endowed with an invariant inner product given
by
\begin{equation}\label{inp}
\la(X,\hat{X}),(Y,\hat{Y})\ra=\la X, Y\ra+\la\hat{X},\hat{Y}\ra,
\end{equation}
where
\begin{equation}\label{}
\la A, B\ra=\int \res (A B)\,\od x \in \cA/\p(\cA)
\end{equation}
for any $A, B\in\cD^\mp$.

Elements of $\cA/\p(\cA)$,  written in the form $\int f\,\od x$ with
$f\in\cA$, are called formal functionals on $\fD$ (viewed as an
infinite-dimensional manifold). In this paper we consider only
formal functionals whose variational gradients belong to $\fD$; by
the variational gradient of a formal functional $F$ at
$\mathbf{A}\in\fD$, it means a pair of pseudo-differential
operators, denoted as $\dt F/\dt\mathbf{A}$, such that $\dt F=\la
\dt F/\dt\mathbf{A},\dt\mathbf{A}\ra$. Note that the variational
gradient of a functional is determined up to some kernel part, and
usually it is not easily written down explicitly.

The algebra $\fD$ is naturally a Lie algebra, whose Lie bracket is
the commutator. On the Lie algebra $\fD$, it follows from a general
result in \cite{Wu} that there is an $R$-matrix defined by (cf.
\cite{Ca, BS})
\begin{equation}\label{Rmatrix}
R(X,\hat{X})=(X_+ -X_- -2\hat{X}_-,\hat{X}_+ -\hat{X}_- +2 X_+),
\end{equation}
and this $R$-matrix solves the modified Yang--Baxter equation.
Moreover, it can be checked
\begin{align*}
\la R(X,\hat{X}),(Y,\hat{Y})\ra=&\la(X_+ - X_- -
2\hat{X}_-)Y+(\hat{X}_+ - \hat{X}_- + 2 X_+)\hat{Y}\ra \\
=&\la X(Y_- - Y_+ +2\hat{Y}_-)+\hat{X}(-Y_+ + \hat{Y}_- -
2\hat{Y}_+)\ra \\
=&\la (X,\hat{X}),-R(Y,\hat{Y})\ra;
\end{align*}
namely, $R$ is anti-symmetric with respect to the inner product
\eqref{inp}. {Thus by using the second assertion of
Theorem~\ref{thm-3br}}, we have the following result.
\begin{lem}\label{thm-PoiD}
On the algebra $\fD$ there is a Poisson bracket between formal
functionals defined by
\begin{equation}\label{Poibra}
\{F, H\}(\mathbf{A})=\left\la\frac{\dt F}{\dt\mathbf{A}},
\cP_{\mathbf{A}}\left(\frac{\dt H}{\dt\mathbf{A}}\right)\right\ra,
\quad \mathbf{A}=(A,\hat{A})\in\fD,
\end{equation}
where the Poisson tensor $\cP: T\fD^*\to T\fD$ (the tangent and the
cotangent fibers are identified with $\fD$) is given by
\begin{align}
\cP_{(A,\hat{A})}(X,\hat{X})=& \big(-(A X+\hat{A}\hat{X})_-A+A(X
A+\hat{X}\hat{A})_-, \nn\\
&\quad (A X+\hat{A}\hat{X})_+\hat{A}-\hat{A}(X
A+\hat{X}\hat{A})_+\big). \label{PAX}
\end{align}
\end{lem}

\subsection{Hamiltonian representation for the extended KP hierarchy}

Let us start to derive Hamiltonian structures for the extended KP
hierarchy \eqref{Ltth}--\eqref{Lhtth}, by reducing the Poisson
bracket \eqref{Poibra} to suitable submanifolds of $\fD$.

Given operators \eqref{P} and \eqref{Ph}, for any positive integer
$m$ we let
\begin{equation}\label{bA}
\mathbf{A}=(A,\hat{A})=(P^{m}, \hat{P}).
\end{equation}
More explicitly, one has
\begin{align}
&A=(\p-\vp)^m+m\, \vp(\p-\vp)^{m-1}+\sum_{i\le m-2}v_{i} (\p-\vp)^i,
\label{A}
\\
&\hat{A}=(\p-\vp)^{-1}\rho +\sum_{i\ge0}\hat{u}_{i} (\p-\vp)^i.
\label{Ah}
\end{align}
Conversely, suppose $A$ and $\hat{A}$ are given as above, then
$P=A^{1/m}$ and $\hat{P}=\hat{A}$ of the form \eqref{P}--\eqref{Ph}
are determined uniquely. Observe that the unknown functions
$(v_{m-2}, v_{m-3}, \dots, \vp, \rho, \hat{u}_{0}, \hat{u}_1,
\dots)$ above and $(u_{1}, u_{2}, \dots, \vp, \rho, \hat{u}_{0},
\hat{u}_1, \dots)$  in \eqref{P}--\eqref{Ph} are up to a Miura-type
transformation. Thus the extended KP hierarchy
\eqref{Ltth}--\eqref{Lhtth} can be represented equivalently as
\begin{align}\label{At}
&\frac{\p\mathbf{A}}{\p t_k}=\left[\left((P^k)_+,(P^k)_+\right),(A,
\hat{A})\right],
\\
&\frac{\p\mathbf{A}}{\p\hat{t}_k}=\left[\left(-(\hat{P}^k)_-,-(\hat{P}^k)_-\right),(A,
\hat{A})\right]. \label{Ath}
\end{align}

For $k\in\Z$ we introduce the following notations based on the
$\vp$-expansion for pseudo-differential operators:
\begin{align}\label{Dlowerk}
\left(\cD_\vp^\mp\right)_{\le k}=&\set{\sum_{i\le k}f_k(\p-\vp)^i\in
\cD^\mp\mid f_i\in\cA}, \\
\left(\cD_\vp^\mp\right)_{\ge k}=&\set{\sum_{i\ge k}f_k(\p-\vp)^i\in
\cD^\mp\mid f_i\in\cA}.
\end{align}
Clearly $\cD^\mp=\left(\cD_\vp^\mp\right)_{\le
k}\oplus\left(\cD_\vp^\mp\right)_{\ge k+1}$.

All operators of the form \eqref{bA} compose a subset of $\fD$ as
\begin{equation}\label{eqU}
\cU_m=\left(\p^m+\left(\cD_\vp^-\right)_{\le m-2}\right) \times
\left(\left(\cD_\vp^+\right)_{\ge 0}\oplus \cM\right),
\end{equation}
where
\[
\cM=\{(\p-\vp)^{-1}\rho \mid\vp,\rho\in\cA; \vp,\rho\ne0\}.
\]
Here $\cM$ is considered as a $2$-dimensional manifold with local
coordinate $(\vp,\rho)$. This manifold has tangent spaces
\begin{equation}
T_{\vp,\rho} \cM=\{(\p-\vp)^{-1}a(\p-\vp)^{-1}\rho + (\p-\vp)^{-1}b
\mid a,b\in\cA\},
\end{equation}
while the cotangent spaces are
\begin{equation}
T^*_{\vp,\rho}\cM=\cA\oplus\cA(\p-\vp).
\end{equation}

For the the subset $\cU_m$, its tangent bundle $T\cU_m$ is composed
by the following fibers
\begin{equation}\label{tanU}
T_{\mathbf{A}}\cU_m=(\cD^-_\vp)_{\le
m-2}\times\big((\cD^+_\vp)_{\ge0}\oplus T_{\vp,\rho} \cM \big).
\end{equation}
The cotangent bundle $T^*\cU_m$  has fibers (dual with the tangent
spaces)
\begin{equation}\label{ctanU}
T_{\mathbf{A}}^*\cU_m=\, _{\ge
-m+1}\!(\cD^-)\times\big((\cD^+_\vp)_{\le-1}\oplus T_{\vp,\rho}^*\cM
\big),
\end{equation}
where we have used notations (cf. \eqref{Dlowerk})
\begin{align}\label{}
 _{\ge k}\!\left(\cD_\vp^\mp\right)=\set{\sum_{i\ge k}(\p-\vp)^i
f_k\in \cD^\mp\mid f_i\in\cA}.
\end{align}

\begin{lem}\label{lem-Poicoset}
On the subset
 $\cU_m$ consisting of operators of the form
 \eqref{bA}, there is a Poisson tensor $\cP^{\mathrm{red}}: T^*\cU_m\to T\cU_m$ defined by
\begin{align}
\cP_{\mathbf{A}}^{\mathrm{red}}(X,\hat{X})=& \big(-(A
X+\hat{A}\hat{X})_-A+A(X
A+\hat{X}\hat{A})_-, \nn\\
&\quad (A X+\hat{A}\hat{X})_+\hat{A}-\hat{A}(X
A+\hat{X}\hat{A})_+\big) \nn\\
&+\frac{1}{m}([f,A],[f,\hat{A}]), \label{PoiU}
\end{align}
where $\mathbf{A}=(A,\hat{A})\in\cU_m$ and
\begin{equation}\label{intf}
f=\p^{-1}\left(\res([X,A]+[\hat{X},\hat{A}])\right)\in\cA
\end{equation}
(note that the residue of a commutator of two pseudo-differential
operators is always a total derivative in $x$).
\end{lem}
\begin{prf}
We recall the Poisson bracket \eqref{Poibra} on $\fD$, and want to
reduce it onto the subset $\cU_m$. To this end, let us consider the
following decompositions of subspaces:
\[
\fD=T_{\mathbf{A}}\cU_m\oplus \cV_{\mathbf{A}}=
T_{\mathbf{A}}^*\cU_m\oplus\cV^*_{\mathbf{A}},
\]
where
\begin{align*}\label{}
&\cV_{\mathbf{A}}=(\cD^-_\vp)_{\ge m-1}\times ((\cD^+_\vp)_{\le-1}/T_{\vp,\rho}\cM),  \\
&
\cV^*_{\mathbf{A}}=(\cD^-_\vp)_{\le-m}\times\left(\frac{1}{\rho}(\p-\vp)(\cD^+_\vp)_{\ge0}(\p-\vp)
\right).
\end{align*}
The Poisson tensor \eqref{PAX} can be written as
\begin{equation*}\label{}
\cP_{\mathbf{A}}=\left(
              \begin{array}{cc}
                \cP_{\mathbf{A}}^{\cU\cU} & \cP_{\mathbf{A}}^{\cU\cV} \\
                \cP_{\mathbf{A}}^{\cV\cU} & \cP_{\mathbf{A}}^{\cV\cV} \\
              \end{array}
            \right): T_{\mathbf{A}}^*\cU_m\oplus \cV^*_{\mathbf{A}}\to
            T_{\mathbf{A}}\cU_m\oplus\cV_{\mathbf{A}}.
\end{equation*}
More exactly, given $(X,\hat{X})\in\cV_{\mathbf{A}}^*$ with
\[
X=X_{-m}(\p-\vp)^{-m}+X_{-m-1}(\p-\vp)^{-m-1}+\dots, \quad
X_{i}\in\cA,
\]
 one has $(A X)_+=(X A)_+=X_{-m}$ and $(\hat{A}
\hat{X})_-=(\hat{X} \hat{A})_-=0$. Hence
\begin{align}
\cP_\mathbf{A}^{\cU\cV} (X,\hat{X})=&\big(-(A X- X_{-m})A+A(X
A-X_{-m}), \nn\\
&\left.(X_{-m}+\hat{A}\hat{X})\hat{A}-\hat{A}(X_{-m}+\hat{X}\hat{A})\big)\right|_{T_{\mathbf{A}}\cU_m}
\nn\\
=&\left.\left([X_{-m},A],[X_{-m},\hat{A}]\right)\right|_{T_{\mathbf{A}}\cU_m}
\nn\\
=&\left([X_{-m},A]+m\,\p(X_{-m})(\p-\vp)^{m-1},[X_{-m},\hat{A}]\right),\label{PUV}
\\
\cP_\mathbf{A}^{\cV\cV}
(X,\hat{X})=&\left.\left([X_{-m},A],[X_{-m},\hat{A}]\right)\right|_{\cV_\mathbf{A}}
\nn\\
=&\left(-m\,\p(X_{-m})(\p-\vp)^{m-1},0\right),
\end{align}
where we have used the fact
\begin{align}\label{}
[X_{-m},\hat{A}]_-=&X_{-m}(\p-\vp)^{-1}\rho-(\p-\vp)^{-1}\rho X_{-m}
\nn
\\
=&(\p-\vp)^{-1}\p(X_{-m})(\p-\vp)^{-1}\rho\in T_{\vp,\rho}\cM.
\label{XAm}
\end{align}
On the other hand, for  $(X,\hat{X})\in T_{\mathbf{A}}^*\cU_m$, one
has
\begin{align}\label{}
\cP_\mathbf{A}^{\cV\cU} (X,\hat{X})=&\big(-\res(A
X+\hat{A}\hat{X})\cdot (\p-\vp)^{m-1}+(\p-\vp)^{m-1} \res(X
A+\hat{X}\hat{A}), \nn\\
&\left.(A X+\hat{A}\hat{X})_+(\p-\vp)^{-1}\rho-(\p-\vp)^{-1}\rho(X
A+\hat{X}\hat{A})_+\big)\right|_{\cV_\mathbf{A}}
\nn\\
=&\big(\res(X A+\hat{X}\hat{A}-A
X-\hat{A}\hat{X})\cdot(\p-\vp)^{m-1},  \nn\\
&\left. a
(\p-\vp)^{-1}\rho - (\p-\vp)^{-1} b\big)\right|_{\cV_\mathbf{A}} \nn\\
= &\big(\p(f)\cdot(\p-\vp)^{m-1},0\big), \label{PVU}
\end{align}
where
\[
a=\res( (A X+\hat{A}\hat{X})(\p-\vp)^{-1}), \quad b=
\res((\p-\vp)^{-1}\rho(X A+\hat{X}\hat{A}) ),
\]
$f$ is given in \eqref{intf}, and the last equality in \eqref{PVU}
holds for the same reason as \eqref{XAm}.

According to \eqref{PUV}--\eqref{PVU}, the following Dirac reduction
(see, for example, \cite{MR}) from $\fD$ to $\cU_{m}$ is feasible:
\[
\cP_\mathbf{A}^{\mathrm{red}}=\cP_\mathbf{A}^{\cU\cU}-
\cP_\mathbf{A}^{\cU\cV}\circ\left(\cP_\mathbf{A}^{\cV\cV}\right)^{-1}\circ
\cP_\mathbf{A}^{\cV\cU},
\]
that is, for $(X,\hat{X})\in T_{\mathbf{A}}^*\cU_m$,
\begin{align}\label{}
\cP_\mathbf{A}^{\mathrm{red}} (X,\hat{X})=&\big(-(A
X+\hat{A}\hat{X})_- A+A(X
A+\hat{X}\hat{A})_-, \nn\\
&(A X+\hat{A}\hat{X})_+\hat{A}-\hat{A}(X
A+\hat{X}\hat{A})_+\big)\nn\\
&-\cP_\mathbf{A}^{\cV\cU} (X,\hat{X})
+\frac{1}{m}\left([f,A]+m\,\p(f)(\p-\vp)^{m-1},[f,\hat{A}]\right)
\nn\\
=&\big(-(A X+\hat{A}\hat{X})_- A+A(X
A+\hat{X}\hat{A})_-, \nn\\
&(A X+\hat{A}\hat{X})_+\hat{A}-\hat{A}(X
A+\hat{X}\hat{A})_+\big) \nn\\
&+\frac{1}{m}\left([f,A],[f,\hat{A}]\right). \label{Pred}
\end{align}
The lemma is proved.
\end{prf}

By now we have obtained a Poisson tensor $\cP^{\mathrm{red}}$ on
$\cU_m$. The following shift transformation
\begin{equation}
\mathscr{S}: (A,\hat{A})\mapsto (A+s,\hat{A}+s),
\end{equation}
where $s$ is a parameter, induces a push-forward of the Poisson
tensor $\cP^{\mathrm{red}}$ as
\[
\mathscr{S}_*\cP^{\mathrm{red}}=\cP_2-s\,\cP_1+s^2\,\cP_0.
\]
By a straightforward calculation, we have $\cP_0=0$, and conclude
the following lemma.
\begin{lem} \label{thm-poi12}
On the subset $\cU_m$ of $\cD_\varphi$ there are two compatible
Poisson tensors defined as follows:
\begin{align}\label{Poi1}
\cP_1(X,\hat{X})=& \big(-[X_- +\hat{X}_-,A]+[X,A]_- +
[\hat{X},\hat{A}]_- \nn\\
& \quad  [X_+ +\hat{X}_+,A]-[X,A]_+ -
[\hat{X},\hat{A}]_+\big), \\
\cP_2(X,\hat{X})=& \big(-(A X+\hat{A}\hat{X})_-A+A(X
A+\hat{X}\hat{A})_-, \nn\\&\quad (A
X+\hat{A}\hat{X})_+\hat{A}-\hat{A}(X A+\hat{X}\hat{A})_+\big) \nn\\
&+\frac{1}{m}\left([f,A],[f,\hat{A}]\right) \label{Poi2}
\end{align}
with $(X,\hat{X})\in T^*_{\mathbf{A}}\cU_m$ at any point
$\mathbf{A}=(A,\hat{A})\in\cU_m$, and $f$ given in \eqref{intf}.
\end{lem}

Finally, let us represent the extended KP hierarchy into a
bi-Hamiltonian form.
\begin{thm}\label{thm-PoieKP}
For any positive integer $m$, let $\{~,\,\}_{1,2}^m$ be the Poisson
brackets on $\cU_m$ given by the tensors $\cP_{1,2}$ in
\eqref{Poi1}--\eqref{Poi2}. The extended KP hierarchy
\eqref{Ltth}--\eqref{Lhtth} can be represented as
\begin{align}\label{Ham2BKP}
&\frac{\p F}{\p t_k}=\{F, H_{k+m}\}^m_1=\{F, H_{k}\}^m_2, \\
&\frac{\p F}{\p \hat{t}_k}=\{F, \hat{H}_{k+1}\}^m_1=\{F,
\hat{H}_{k}\}^m_2 \label{HameKP}
\end{align}
with $k=1,2,3,\dots$ and Hamiltonians
\begin{align}\label{HHhat}
H_{k}=\frac{m}{k}\int\res P^k\,\od x, \quad
\hat{H}_{k}=\frac{1}{k}\int\res \hat{P}^k\,\od x.
\end{align}
\end{thm}
\begin{prf}
The proof is similar to that of Theorem~4.6 in \cite{Wu}. Recalling
$\mathbf{A}=(A,\hat{A})=(P^m,\hat{P})$, since
\[
\dt H_k=\la(P^{k-m},0),\dt\mathbf{A}\ra, \quad \dt \hat{H}_k=\la(0,
\hat{P}^{k-1}),\dt\mathbf{A}\ra,
\]
then the gradients of the Hamiltonian functionals are
\begin{equation}\label{}
\frac{\dt H_k}{\dt\mathbf{A}}=(P^{k-m},0), \quad \frac{\dt
\hat{H}_k}{\dt\mathbf{A}}=(0, \hat{P}^{k-1})
\end{equation}
up to a kernel part in $\cV_\mathbf{A}^*$. Note that such a kernel
part does not change the following Hamiltonian vector fields
\begin{align}\label{}
\cP_2\left(\frac{\dt H_{k}}{\dt\mathbf{A}}\right)=&(-(P^k)_-A+A
(P^k)_-, (P^k)_+\hat{A}-\hat{A}(P^k)_+)
\nn\\
=&[((P^k)_+,(P^k)_+),(A,\hat{A})],
\\
\cP_2\left(\frac{\dt
\hat{H}_{k}}{\dt\mathbf{A}}\right)=&(-(\hat{P}^k)_-A+A
(\hat{P}^k)_-,
(\hat{P}^k)_+\hat{A}-\hat{A}(\hat{P}^k)_+)\nn\\
=&[((-\hat{P}^k)_-,(-\hat{P}^k)_-),(A,\hat{A})].
\end{align}
Namely, by virtue of \eqref{At}--\eqref{Ath} we have
\begin{align}\label{}
&\frac{\p\mathbf{A}}{\p t_k}=\cP_2\left(\frac{\dt
H_{k}}{\dt\mathbf{A}}\right), \quad
\frac{\p\mathbf{A}}{\p\hat{t}_k}=\cP_2\left(\frac{\dt
\hat{H}_{k}}{\dt\mathbf{A}}\right).
\end{align}
In the same way, it can be checked
\begin{align}\label{}
&\frac{\p\mathbf{A}}{\p t_k}=\cP_1\left(\frac{\dt
H_{k+m}}{\dt\mathbf{A}}\right), \quad
\frac{\p\mathbf{A}}{\p\hat{t}_k}=\cP_1\left(\frac{\dt
\hat{H}_{k+1}}{\dt\mathbf{A}}\right).
\end{align}
Therefore the theorem is proved.
\end{prf}

In the above theorem we obtain a family of bi-Hamiltonian structures
with index $m$ for the extended KP hierarchy. It is natural to ask
whether there is a certain class of $(m,n)$-parameterized
bi-Hamiltonian structures (corresponding to $(P^m, \hat{P}^n)$ )
such like the case for the Toda lattice hierarchy studied in
\cite{Wu}. The answer is unknown yet, for the reason that, when
$n\ge2$, it is complicated to do calculations in tangent/cotangent
spaces of the submanifold consisting of pseudo-differential
operators of the form $\hat{P}^n$. However, such kind of difficulty
does not appear in the dispersionless limit case, for which we will
derive a two-parameter family of bi-Hamiltonian structures, see
Proposition~\ref{thm-hamd2kp} below.

\subsection{Relation to the constrained KP hierarchy}

Let us consider a natural reduction of the extended KP hierarchy
\eqref{Ltth}--\eqref{Lhtth}, as well as its Hamiltonian structures.

Recall the pseudo-differential operators \eqref{A}--\eqref{Ah}. Now
assume $A=\hat{A}$, each equal to
\begin{equation}\label{L}
L=(\p-\vp)^m+m\vp(\p-\vp)^{m-1}+\sum_{i=0}^{m-2}v_i(\p-\vp)^i+(\p-\vp)^{-1}\rho.
\end{equation}
Under this assumption, the extended KP hierarchy
\eqref{At}--\eqref{Ath} is reduced to
\begin{equation}\label{Lt}
\frac{\p L}{\p t_k}=[(L^{k/m})_+,L], \quad k=1,2,3,\dots,
\end{equation}
where $L^{1/m}=P$ takes the form \eqref{P}.

\begin{prp}\label{thm-redham}
The hierarchy \eqref{Lt} can be represented in a bi-Hamiltonian form
as
\begin{align}\label{}
\frac{\p L}{\p t_k}=\cP_1\left(\frac{\dt H_{k+m}}{\dt
L}\right)=\cP_2\left(\frac{\dt H_{k}}{\dt L}\right), \quad
k=1,2,3,\dots,
\end{align}
where $\cP_{1,2}$ are Poisson tensors given by
\begin{align}\label{P1Y}
&\cP_1(Y)=-[Y_-,L]+[Y,L]_-, \\
&\cP_2(Y)=-(L Y)_-L+L(Y L)_-+\frac{1}{m}[g,L] \label{P2Y}
\end{align}
with $g=\p^{-1}\left(\res[Y, L]\right)$,  and the Hamiltonian
functionals are
\begin{align}\label{HHhat}
H_{k}=\frac{m}{k}\int\res L^{k/m}\,\od x.
\end{align}
\end{prp}
\begin{prf}
Let $\cA_{(A,\hat{A})}$ be the algebra of formal differential
polynomials generated by the unknown functions in $(A,\hat{A})$
given in \eqref{A}--\eqref{Ah}, and $\cA_L$ be the differential
algebra generated by the unknown functions in \eqref{L}. For any
functional $F\in\cA_L/\p(\cA_L)$, the embedding
$\cA_L\hookrightarrow\cA_{(A,\hat{A})}$ due to $L=A=\hat{A}$ implies
the following relation of variational gradients
\[
\frac{\dt F}{\dt L}=\frac{\dt F}{\dt A}+\frac{\dt F}{\dt \hat{A}}.
\]
Accordingly, it is straightforward to check that the Poisson tensors
\eqref{Poi1}--\eqref{Poi2} are reduced to \eqref{P1Y}--\eqref{P2Y}.
Therefore the proposition is proved.
\end{prf}

Observe that the bi-Hamiltonian structure in the above proposition
is equivalent to that obtained by Oevel and Strampp \cite{OS} and by
Cheng \cite{CY}. The central invariants for the bi-Hamiltonian
structure were calculated recently by Liu, Zhang and one of the
authors \cite{LZZ}, based on an explicit formula for the variational
gradient ${\dt F}/{\dt L}$.

\begin{exa}
When $m=1$, we have
\begin{equation}\label{}
L=(\p-\vp)+\vp+(\p-\vp)^{-1}\rho=\p+(\p-\vp)^{-1}\rho.
\end{equation}
The hierarchy \eqref{Lt} now reads
\begin{equation}\label{Lt1}
\frac{\p L}{\p t_k}=[(L^{k})_+,L], \quad k=1,2,3,\dots.
\end{equation}
One has $\p/\p t_1=\p/\p x$, and according to
\eqref{phit}--\eqref{rhot},
\begin{equation}\label{rhophit2}
\frac{\p\vp}{\p t_2}=\p_x(\vp^2+2\rho+\vp_x), \quad \frac{\p\rho}{\p
t_2}=\p_x(2\vp\rho-\rho_x).
\end{equation}
Here the subscript $x$ means the derivative with respect to it.

If we perform the following transformation of variables:
\begin{equation}\label{tranqr}
\vp=\frac{q_x}{q}, \quad \rho=q\,r,
\end{equation}
then the equations \eqref{rhophit2} are converted to
\begin{equation}\label{qrt2}
\frac{\p q}{\p t_2}=2 q^2 r+q_{xx}, \quad \frac{\p r}{\p t_2}=-2
q\,r^2-r_{xx},
\end{equation}
which form the nonlinear Schr\"{o}dinger equation. We also note that
the hierarchy \eqref{Lt1} together with certain extra flows composes
the so-called extended nonlinear Schr\"{o}dinger hierarchy in
\cite{CDZ}, which is equivalent to the extended Toda hierarchy up to
a transformation of variables. An interesting question is whether
the extended KP hierarchy and the Toda lattice hierarchy are related
in a similar way; we plan to study it somewhere else.
\end{exa}

\begin{exa}
When $m=2$, we have
\begin{equation}\label{}
L=(\p-\vp)^2+2\vp(\p-\vp)+u+(\p-\vp)^{-1}\rho,
\end{equation}
and the hierarchy \eqref{Lt} is the so-called Yajima--Oikawa
hierarchy  \cite{CY,YO}. The first nontrivial equations are
\begin{equation}\label{rhophit3}
\frac{\p\vp}{\p t_2}=u_x, \quad \frac{\p\rho}{\p
t_2}=\p_x(2\vp\rho-\rho_x), \quad \frac{\p u}{\p t_2}=2\rho_x+2\vp
u_x+u_{x x}.
\end{equation}
In fact, these equations are recast to equations (2.20) in \cite{CY}
via the following transformation:
\begin{equation}\label{}
\vp=\frac{q_x}{q}, \quad \rho=q\,r,\quad u=u_1+\frac{q_{x x}}{q}.
\end{equation}
\end{exa}

\begin{rmk}
Let $A$ and $B$ be two differential operators of the form
\begin{align}\label{}
&A=\p^{m+n}+a_{m+n-1}\p^{m+n-1}+\dots+a_{1}\p+a_{0}, \\
&B=\p^n+b_{n-1}\p^{n-1}+\dots+b_{1}\p+b_0,
\end{align}
and $L=B^{-1}A$ with $B^{-1}=\p^{-n}-b_{n-1}\p^{-n-1}+\dots$. Recall
that a version of constrained KP hierarchy suggested by Aratyn et al
\cite{ANP} and by Bonora et al \cite{BX} (see also \cite{Di}) is the
collection of the following evolutionary equations
\begin{equation}\label{cKP}
\frac{\p A}{\p t_k}=-\left(A(L^{k/m})_+ A^{-1}\right)_-A, \quad
\frac{\p B}{\p t_k}=-\left(B(L^{k/m})_+ B^{-1}\right)_-B,
\end{equation}
where $k=1,2,3,\dots$. In particular, one has
\begin{align}\label{}
\frac{\p L}{\p t_k}=&-B^{-1}\frac{\p B}{\p
t_k}B^{-1}A+B^{-1}\frac{\p A}{\p t_k} \nn\\
=&B^{-1}\left(B(L^{k/m})_+ B^{-1}-A(L^{k/m})_+
A^{-1}\right)_-A \nn\\
=&B^{-1}\left(B\left[(L^{k/m})_+, B^{-1}A\right] A^{-1}\right)_-A
\nn\\
=&\left[(L^{k/m})_+, L\right],
\end{align}
of which the last equality holds for the fact
\[
\left[(L^{k/m})_+, L\right]=-\left[(L^{k/m})_-, L\right]
\]
having order less than $m$. Hence the hierarchy \eqref{Lt} can be
regarded as a reduction of the constrained KP hierarchy \eqref{cKP}
under the assumption
\begin{equation}\label{}
A=\p^{m}+a_{m-2}\p^{m-2}+\dots+a_{1}\p+a_{0}, \quad B=\p-\vp.
\end{equation}
From this point of view, the second Hamiltonian structure
\eqref{P2Y} is just a reduction of the Hamiltonian structure
obtained by Dickey \cite{Di} for the constrained KP hierarchy
\eqref{cKP}.
\end{rmk}

\subsection{Dispersionless limit}

For the extended KP hierarchy, let us study its dispersionless limit
and the corresponding Hamiltonian structures.

Given a function $\vp\in\cA_1$ as before, we introduce the following
two algebras:
\begin{align}\label{}
&\cH_\vp^-=\cA((\,(z-\vp)^{-1}))=\set{\sum_{i\le k}f_i(z-\vp)^i\mid
f_i\in\cA, k\in\Z},\\
&\cH_\vp^+=\cA((z-\vp))=\set{\sum_{i\ge k}f_i(z-\vp)^i\mid
f_i\in\cA, k\in\Z}.
\end{align}
They consist of Laurent series in $z-\vp$ defined outside or inside
a loop $\Gm_\vp$ surrounding $z=\vp$ on the complex plane
respectively. Each algebra $\cH_\vp^\pm$ is endowed with a Lie
bracket given by
\begin{equation}\label{}
[a,b]=\frac{\p a}{\p z}\frac{\p b}{\p x}-\frac{\p b}{\p z}\frac{\p
a}{\p x}.
\end{equation}
Moreover, on each  $\cH_\vp^\pm$ there is an inner product
\begin{equation}\label{}
\la a, b\ra=\la a\,b\ra, \quad \la
a\ra=\frac1{2\pi\sqrt{-1}}\oint_{S^1}\oint_{\Gm_\vp}a(z)\,\od z\,\od
x,
\end{equation}
which is invariant with respect to the above Lie bracket.

Introduce two series
\begin{align}\label{}
&p(z)=(z-\vp)+\vp+\sum_{i\ge1}u_i\,(z-\vp)^{-i}\in\cH_\vp^-, \\
&\hat{p}(z)=\sum_{i\ge-1}\hat{u}_i\,(z-\vp)^{i}\in\cH_\vp^+
\end{align}
with $\hat{u}_{-1}=\rho\ne0$. The dispersionless extended KP
hierarchy (or the Whitham hierarchy of genus zero with one marked
point \cite{SB, TT}, without the logarithm flow) is defined by
\begin{equation}\label{deKP}
\frac{\p \dot{p}(z)}{\p t_k}=[(p(z)^k)_+,\dot{p}(z)], \quad \frac{\p
\dot{p}(z)}{\p \hat{t}_k}=[-(\hat{p}(z)^k)_-,\dot{p}(z)],
\end{equation}
where $\dot{p}(z)=p(z), \hat{p}(z)$, and $k=1,2,3,\dots$. Here the
subscripts ``$\pm$'' mean to take the nonnegative and negative
parts, respectively, of a series in $z-\vp$.

Let us derive Hamiltonian structures for the dispersionless
hierarchy \eqref{deKP}. To this end, we consider the Lie algebra
$\mathfrak{H}_\vp=\cH_\vp^-\times\cH_\vp^+$ whose Lie bracket is
defined diagonally. On $\mathfrak{H}_\vp$ it is equipped with an
inner product as
\[
\la(a,\hat{a}),(b,\hat{b})\ra=\la a,b\ra+\la\hat{a},\hat{b}\ra,
\quad (a,\hat{a}),~(b,\hat{b})\in \mathfrak{H}_\vp.
\]
Clearly, a map $R$ of the form \eqref{Rmatrix} defines an $R$-matrix
that solves the modified Yang--Baxter equation on
$\mathfrak{H}_\vp$.

{According to Theorem~\ref{thm-3br}, one has the following Poisson
bracket }
\begin{equation}\label{pbr}
\{F,H\}(\bm{a})=\frac{1}{2}\left(\left\la\left[\bm{a},\frac{\dt
F}{\dt\bm{a}}\right],R\left(\bm{a}\frac{\dt
H}{\dt\bm{a}}\right)\right\ra-\left\la\left[\bm{a},\frac{\dt
H}{\dt\bm{a}}\right],R\left(\bm{a}\frac{\dt
F}{\dt\bm{a}}\right)\right\ra\right),
\end{equation}
where $F, H$ are arbitrary functionals depending on
$\bm{a}\in\mathfrak{H}_\vp$, with variational gradients $\dt
F/\dt\bm{a}, \dt H/\dt\bm{a}$ respectively (the definition of
variational gradient is almost the same as before, so it is omitted
here).


Given two arbitrary positive integers $m$ and $n$, let
\begin{equation}\label{}
\bm{a}(z)=(a(z),\hat{a}(z))=(p(z)^{m},\hat{p}(z)^{n}).
\end{equation}
All such series form a subset of $\mathfrak{H}_\vp$ as
\begin{equation}\label{}
U_{m,n}=\set{\left(z^m+\sum_{i\le m-2}v_i\,(z-\vp)^{i},\sum_{i\ge
-n}\hat{v}_i\,(z-\vp)^{i}\right)\in\cH_\vp^-\times\cH_\vp^+}.
\end{equation}
\begin{lem}
On $U_{m,n}$ there are two compatible Poisson brackets
$\set{~,~}^{m,n}_\nu$ ($\nu=1,2$) given by the following Poisson
tensors:
\begin{align}
\cP_1(X ,\hat{X} ) =&\big([X ,a ]_-+[\hat{X} ,\hat{a} ]_- -[X _-
+\hat{X}_-,a ], \nn\\
&\quad~ -[X , a ]_+ -[\hat{X} ,\hat{a} ]_+ +[X_+ +\hat{X}_+,\hat{a}
]\big),
\label{dP1} \\
\cP_2(X  ,\hat{X}  ) =&\big(([X ,a ]_-+[\hat{X} ,\hat{a} ]_-)a -[(X
a +\hat{X} \hat{a} )_-,a ], \nn\\
&\quad~ -([X , a ]_+ +[\hat{X} ,\hat{a} ]_+)\hat{a}
+[(X  a +\hat{X} \hat{a} )_+,\hat{a} ]\big) \nn\\
&+\frac{1}{m}\left([f,a],[f,\hat{a}]\right), \label{dP2}
\end{align}
where
\begin{equation}\label{}
f=\p^{-1}\left(\res_{z=\vp} ([X,a]+[\hat{X},\hat{a}]) \right), \quad
(X ,\hat{X})\in T^*_{(a,\hat{a})} U_{m,n}.
\end{equation}
\end{lem}
\begin{prf}
The proof is similar with (and easier than) that of
Lemma~\ref{lem-Poicoset}, with the method of Dirac reduction, so we
omit the details here.
\end{prf}

Furthermore, in the same way as for Theorem~\ref{thm-PoieKP} we
arrive at
\begin{prp} \label{thm-hamd2kp}
For any positive integers $m$ and $n$, the dispersionless extended
KP hierarchy \eqref{deKP} can be represented in a bi-Hamiltonian
form as
\begin{align}\label{Hamd2BKP}
&\frac{\p F}{\p t_k}=\{F, H_{k+m}\}^{m,n}_1=\{F, H_{k}\}^{m,n}_2,
\\
&\frac{\p F}{\p \hat{t}_k}=\{F, \hat{H}_{k+n}\}^{m,n}_1=\{F,
\hat{H}_{k}\}^{m,n}_2 \label{Hamd2BKP2}
\end{align}
with $k=1,2,3,\dots$ and
\begin{equation*}\label{}
H_{k}=\frac{m}{k}\la p(z)^k\ra, \quad \hat{H}_{k}=\frac{n}{k}\la
\hat{p}(z)^k\ra.
\end{equation*}
\end{prp}

Let us consider reductions of the dispersionless extended KP
hierarchy. Assume
\begin{equation}\label{pphl}
p(z)^{m}=\hat{p}(z)^{n}=l(z)
\end{equation}
with
\[
l(z)=(z-\vp)^{m}+m\,\vp(z-\vp)^{m-1}+\sum_{i=-n}^{m-2}v_i\,(z-\vp)^{i},
\]
then the hierarchy \eqref{deKP} is reduced to
\begin{equation}\label{dmn}
\frac{\p l(z)}{\p t_k}=[(p(z)^k)_+,l(z)], \quad \frac{\p l(z)}{\p
\hat{t}_k}=[-(\hat{p}(z)^k)_-,l(z)],
\end{equation}
where $k=1,2,3,\dots$.

With the same method as for Proposition~\ref{thm-redham}, we have
\begin{prp}
Under the constraint \eqref{pphl}, the Poisson structures
\eqref{dP1}--\eqref{dP2} are reduced to
\begin{align}
\cP_1(Y ) =&[Y ,l]_- -[Y_-
,l],  \\
\cP_2(Y) =&[Y ,l ]_-l -[(Y l)_-,l ] +\frac{1}{m}[g,l]
\end{align}
with $g=\p^{-1}\left(\res_{z=\vp}[Y,l]\right)$. They give a
bi-Hamiltonian structure for the reduced hierarchy \eqref{dmn}.
\end{prp}

\section{Concluding remarks}

In this article we have considered the extended KP hierarchy defined
with scalar pseudo-differential operators, whose dispersionless
limit is the Whitham hierarchy with only one marked point. We
clarify how the hierarchy is related to the $2$-component BKP
 and the constrained KP hierarchies. We hope that the
reconstruction of the hierarchy with two types of
pseudo-differential operators, as well as the study of its
Hamiltonian structures, would help to cause more attention to this
topic, especially its possible relation with infinite-dimensional
Frobenius manifolds \cite{CDM, WX, WZ} (see also \cite{Ra, Sz}). We
will study it elsewhere.

By using the density functions of the Hamiltonians \eqref{HHhat},
one has the following  closed $1$-form
\[
\om=\sum_{k=1,2,3,\dots}(\res\,P^k\,\od
t_k+\res\,\hat{P}^k\,\od\hat{t}_k).
\]
Hence given a solution of the extended KP hierarchy
\eqref{Ltth}--\eqref{Lhtth} there locally exists a tau function
$\tau=\tau(t_1,t_2,\dots;\hat{t}_1,\hat{t}_2,\dots)$ such that
\begin{align}\label{tau}
\om=\od(\p_x\log\tau).
\end{align}
In contrast to the KP hierarchy, it is unknown whether the extended
KP hierarchy \eqref{Ltth}--\eqref{Lhtth} could be recast to some
bilinear equation of $\tau$, such like those Hirota equations raised
via boson-fermion correspondence \cite{KvdL, DKJM-KPBKP}. The main
difficulty in doing this is the dependence of the operator $\p-\vp$
on $\bt$, which is essentially different from the cases of KP and
$2$-component BKP hierarchies. We hope to study it in the future; an
answer to this question would help us to understand generalizations
of the KP hierarchy like \cite{CM, KvdL, TT}.

\vskip 0.5truecm \noindent{\bf Acknowledgments.} {The authors thank
Blazej Szablikowski and Dafeng Zuo for referring them to the work
\cite{SB, BS}.} They also thank the anonymous referee for his very
useful comments. This work is partially supported by the National
Natural Science Foundation of China No. 11171176, No. 11222108, No.
11371214 and No. 11401599, and by the Fundamental Research Funds for
the Central Universities No. 15lgpy33.

\end{document}